\documentclass{ecai} 

\usepackage{latexsym}
\usepackage{amssymb}
\usepackage{amsmath}
\usepackage{amsthm}
\usepackage{booktabs}
\usepackage{enumitem}
\usepackage{graphicx}
\usepackage{color}
\usepackage{float}
\usepackage{placeins}

\newcommand{\BibTeX}{B\kern-.05em{\sc i\kern-.025em b}\kern-.08em\TeX}

\begin{document}


\begin{frontmatter}




\title{Emotion-sensitive Explanation Model}


\author[A,B]{\fnms{Christian}~\snm{Sch\"utze}\orcid{0000-0002-8860-0478}\thanks{Corresponding Author. Email: Christian.Schuetze@uni-bielefeld.de}}
\author[A,B]{\fnms{Birte}~\snm{Richter}\orcid{0000-0002-0957-2406
}}
\author[A,B]{\fnms{Britta}~\snm{Wrede}\orcid{0000-0003-1424-472X}} 

\address[A]{Medical Assistance Systems, Medical School OWL}
\address[B]{Center for Cognitive Interaction Technology (CITEC), Bielefeld University}

\begin{abstract}
Explainable AI (XAI) research has traditionally focused on rational users, aiming to improve understanding and reduce cognitive biases. However, emotional factors play a critical role in how explanations are perceived and processed.
Prior work shows that prior and task-generated emotions can negatively impact the understanding of explanation.
Building on these insights, we propose a three-stage model for emotion-sensitive explanation grounding: (1) emotional or epistemic arousal, (2) understanding, and (3) agreement.
This model provides a conceptual basis for developing XAI systems that dynamically adapt explanation strategies to users’ emotional states, ultimately supporting more effective and user-centered decision-making.
\end{abstract}

\end{frontmatter}


\section{Introduction}

Supporting human decision-making has been in the focus of research for decades \cite{schmid2022explainable}. However, the underlying assumption in such endeavors has mostly been that interaction takes place with a rational decision maker who follows purly logical considerations. Thus, support has been intended to (1) provide the human decision maker with relevance information about certain features, and (2) to avoid cognitive biases such as confirmation bias \cite{wang2019designing, battefeld2024revealing}.

Yet, it is well known that human decision-making is heavily influenced by emotions \cite{lerner2015emotion}.
More recently, emotions have been investigated in the context of XAI and decision-making.
However, most research focuses on the analysis of the effects that emotions have on the explanation process or their acceptance. 

In \cite{thommes2024xai}, it was shown that emotions affected how humans react to explanations. Their results indicate that low arousal decreases the need for explanation whereas high arousal increases positive reactions in terms of advice taking towards explanations.

On the other hand, explanations can also affect emotions.
For example, \cite{guerdan2021iccv} found that giving explanations for an easy task yields negative affect whereas advice for a difficult task yielded positive affect.

\cite{richter2025emotion} investigated the influence of both prior and task-generated emotions on explanation retention and understanding in the context of XAI. 
Neither emotion induction nor task-generated emotional reactions were significant predictors of retention. However, features encoding certain individual characteristics—such as gender, current health status, and political orientation—emerged as significant predictors for the recall of explained features. 
These features were more likely to verbally reproduced.
While no significant main effect of the emotion induction condition on retention was found, the effect on understanding was marginally significant. 
These results suggest a potential trend indicating that task-unrelated emotions may influence participants’ comprehension of explanations.
Notably, emotional reactions were significantly negatively associated with explanation understanding, suggesting a possible disruptive effect of emotional intensity on cognitive processing in XAI contexts.

Taking this one step further, \cite{lammert2025} investigated the effect of nudging on explainees reaction. Nudging was applied in order to support emotional debiasing. While nudging was shown to have an effect on the emotion, no significant effect on decision making could be observed.

\section{Literature}

 These results show that emotions occur during explanations and can affect decision-making. However, there is little research in the context of XAI or human-centered XAI with the goal to develop approaches for achieving understanding in emotional explainees through specific emotion-sensitive explanation strategies.

There is a large body of research investigating how cognitive biases, especially confirmation bias, can be addressed in XAI. A recent approach suggests the use of so called \textit{cognitive forcing functions} \cite{buccinca2021hci}. The basic idea is to increase the explainee's cognitive effort by different strategies, ranging from asking the user to make his/her own decision first, before the XAI is presenting its results, over slowing down the decision making process e.g. by difficult to read text or graphics to providing XAI hypotheses only on demand.
Very detailed and context-dependent strategies are proposed by \cite{simkute2021jrt}.

However, there are so far no approaches that address the question of emotion-sensitive explanation from the perspective of understanding. Accordingly, there don't exist modeling approaches for explaining XAI hypotheses to emotional explainees.

\subsection{Grounding approaches for establishing shared understanding}

Grounding has been proposed as a general principle for establishing joint understanding between two interaction partners (Clark, grounding). It is achieved by a speaker presenting a statement or request to be considered by the interaction partner. This partner will then issue a so called ``acceptance'', indicating whether or not s/he has perceived, processed and understood the statement. At this level, s/he can either signal understanding and proceed to a follow-up statement or give the turn back to the speaker, or s/he can initiate a clarification dialog to yield understanding. The statement will then belong to the common ground that has been established in the interaction. 
This principle is the basis for many human-computer interaction frameworks. For example, \cite{brennan1995kbs} present a telephone system that can process spoken commands such as ``Call X''. Based on insights from user studies, they determined that a simple signal for non-understanding was not sufficient as it would leave the user clue-less as to what has gone wrong and how an occurred problem can be fixed. Therefore, they proposed a hierarchy of grounding levels where different, mostly non-verbal signals were assigned to indicate whether this level was reached successfully or not. For example, it would provide a brief tone to indicate it was ready to listen to an utterance, or provide different tones or melodies for parsing and interpreting. When a problem occurred the system would specify, for example, that it could not interpret the sentence indicating for the user that s/he may have used a command that is not in the (current) repertoire of the system. Also, at the higher level, extra grounding loops could be initiated. For example, before starting to call a certain person, the system would ask the user for confirmation to avoid the execution of wrongly interpreted commands.
Thus, in order to achieve shared understanding interaction partners have to go through different levels of grounding, signalling one's own and monitoring the other one's state of understanding.

Our results from previous studies \cite{richter2025arxiv} on the influence of emotions on understanding of explanations indicate that (1) task unrelated emotions can influence the understanding of explanations, (2) explanations can induce emotional arousal, and (3) such task induced arousal may affect understanding in unexpected ways, often depending on the interaction partner's idiosyncratic experiences and representations. 

\begin{figure}[H]
    \centering
    \includegraphics[width=0.5\linewidth]{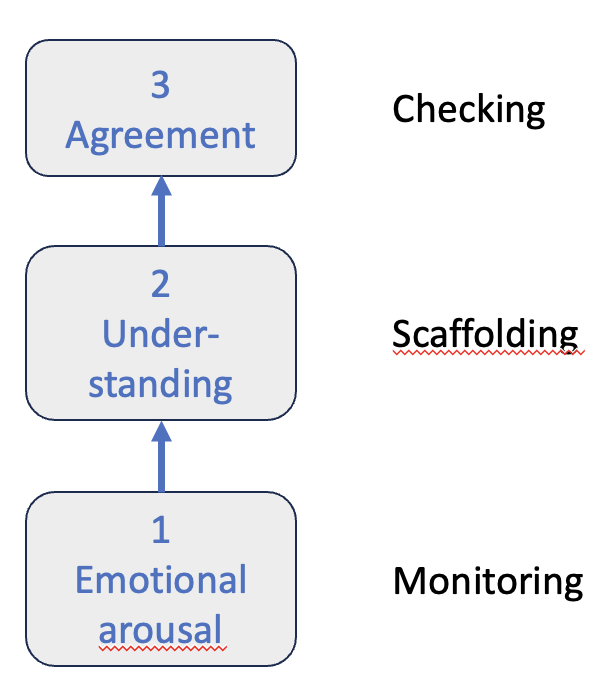}
    \caption{Emotion-sensitive grounding hierarchy for monitoring and scaffolding during feature explanations.}
    \label{fig:A03_emotional-grounding-hierarchy}
\end{figure}

Based on these results, we propose a grounding hierarchy that is sensitive to emotional reactions which may require a specific grounding approach (cf. Fig. \ref{fig:A03_emotional-grounding-hierarchy}).
Thus, at the first level, the explanation system while providing the explanation for a feature, will look out for emotional or epistemic reactions such as irritation or surprise indicating that the explainee may have an issue with the currently explained feature. After detecting a (possibly minimal) reaction, a clarification loop is initiated by asking the user if s/he sees a problem with this explanation, thus starting a clarification sub-dialog. This clarification dialog will contain a range of different explanation strategies, ranging from simple repetition of an argument over rephrasing and contrasting to a change of focus. 
Note, that by reacting to rather subtle features of the interaction partner, the dialog will be possible for mixed initiative. This is an important feature, as the explaining component needs to be able to initiate a dialog when detecting potential misunderstanding.
After the explanation of the feature has been clarified the final step will be to assess whether or not the explainee agrees with this explanation. Note, that the goal is not necessarily that the user agrees to the explanation. Rather, the goal is to provide sufficient information for a well-informed decision. This may include the co-construction of a joint decision, for example, based on a decision taken from a hypothesis of the AI system without a feature that has been critically discussed.

\begin{figure*}[t]
    \centering
    \includegraphics[width=1\linewidth]{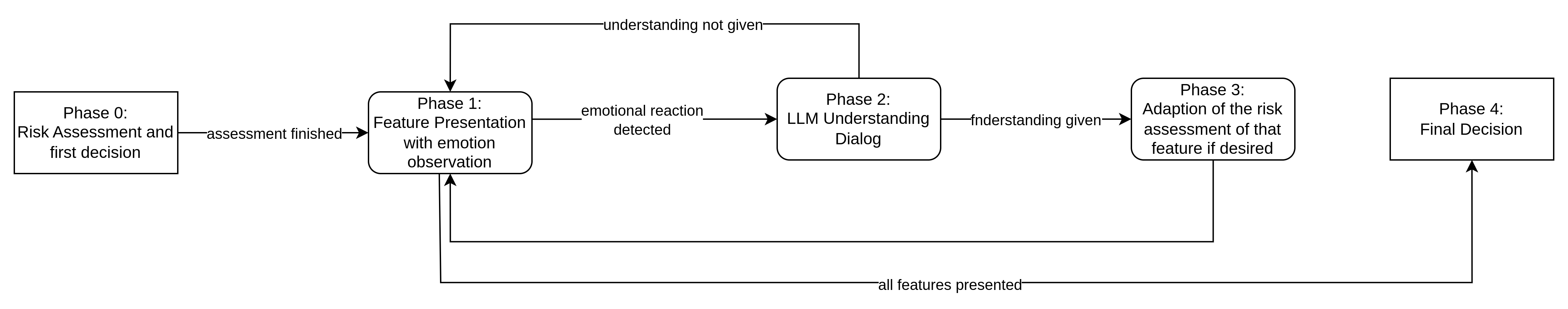}
    \caption{Visualization of the different phases of the emotion-sensitive explanation model.}
    \label{fig:modelPhases}
\end{figure*}

The emotion-sensitive grounding hierarchy is the underlying mechanism of the model for an emotion-sensitive dialog model for decision support in emotional situations.


\section{Model}

Based on results of the influence on decision-making with a Decision Support System (DSS) so far, we have developed an \textbf{emotion-sensitive computational explanation model} that realizes a multimodal interaction with the virtual robot Floka and monitors and scaffolds the explainee according to an \emph{emotional grounding hierarchy}.
The model provides explanations for the features that have led to the system's proposal for selecting a specific risk level one after the other. While providing the initial explanation of a feature (or variable), it monitors the explainee's facial expression and heart rate to detect changes in arousal or emotional expression. This emotional grounding process is based on the grounding hierarchy \cite{brennan1995kbs} with eight grounding steps of their telephone system, ranging from 0 \emph{not attending} to 7 \emph{reporting} (cf. left side of Fig. \ref{fig:A03_emotional-grounding-hierarchy}).

We transfer this hierarchy to the human explainee and extend its notion towards a concept for monitoring and scaffolding the explainee's emotional and epistemic state during the explanation. More specifically, we foresee three emotion-sensitive grounding steps that can encompass different multimodal dialog steps: 
\begin{enumerate}
    \item \textbf{Emotional or epistemic arousal}
    \item \textbf{Understanding}, and
    \item \textbf{Agreement}.
\end{enumerate}

Figure \ref{fig:modelPhases} visualizes the different phases of the emotion-sensitive explanation model.
After the risk assessment and the first assessment (\textit{Phase 0}),  the user’s  \textbf{arousal state} is observed (\textit{Phase 1}) using real-time emotion recognition (via facial expressions detected by EmoNet \cite{gerczuk2021tac}) and physiological indicators such as heart rate variability measured by a smartwatch. 
If a deviation is detected, the system will start to scrutinize the user's  \textbf{understanding} (\textit{Phase 2}) by initiating a dialog regarding the meaning of the feature for the system's risk type classification of the user. 
Once a certain level of understanding has been established, the system moves to assess \textbf{agreement} (\textit{Phase 3}), i.e., whether the user has strong reservations about specific features. 
The underlying assumption of this step is that the training data of an AI system may be biased, outdated, or inappropriate for any reason. For example, gender may be an important feature to classify the user as less risk-oriented. However, this may be based on outdated data. In such cases, the system should provide information about a decision without this information or a counterfactual result.
If no arousal is detected, the system will continue with the explanation of the next feature, see fig \ref{fig:modelPhases}. 
At the end (\textit{Phase 4}), the user makes a final decision.

Note that this approach addresses both the effect of prior, possibly task-unrelated emotions on the user's understanding and the effect that explanations may have on the user's arousal. 


\begin{figure}
    \centering
    \includegraphics[width=1\linewidth]{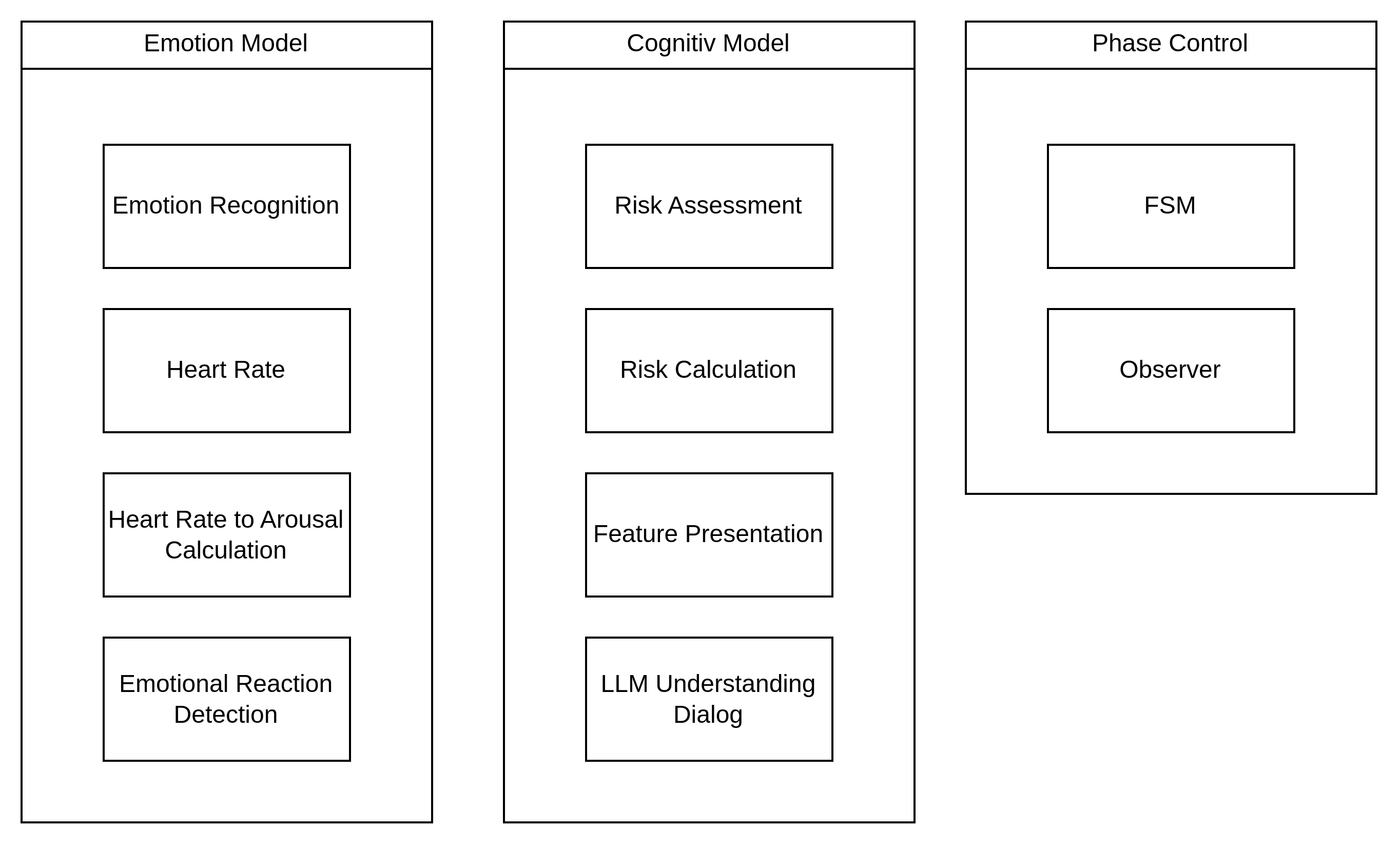}
    \caption{Simplified structural visualization of the emotion-sensitive explanation model.}
    \label{fig:modelStructure}
\end{figure}

Figure \ref{fig:modelStructure} shows the simplified version of the emotion-sensitive explanation model, which is based on three categories: 
\begin{enumerate}
    \item \textbf{Emotional Model} --  Emotion recognition and emotional reaction detection occur in this module. Currently, emotion recognition is done via EmoNet as observer pattern. Therefore, additional arousal sources, e.g., by heart rate and calculation of arousal changes can be added to increase the sensitivity. Emotional reaction detection is implemented via anomaly detection using a rolling z-score (threshold = 2.5) within a 500ms window to capture microexpressions.
    \item \textbf{Cognitive Model } -- Within the cognitive model, risk assessment is performed with different questions and their risk tendencies. Additionally, the presentation of the feature is a fusion of our guided and full transparency strategies \cite{lammert2024fbe}.
    This is achieved by presenting all features while adapting the distribution to strike a balance between risk-averse and risk-tendency features. Additionally, this module hosts the LLM-based understanding dialog, where the user interacts with Floka to reflect on emotionally salient features.
    \item \textbf{Phase Control} -- This component manages the observer’s data flow and orchestrates the transitions between phases. It allows the integration of additional phases if needed and ensures coherent interaction across emotional and cognitive components.
\end{enumerate}


\section{Discussion}

Recent research shows that prior and task-generated emotions can negatively impact the understanding of explanation.
The proposed three-stage model addresses this by monitoring emotional and epistemic states and adapting explanation strategies accordingly. 
The model aligns with prior findings that excessive or insufficient arousal impairs decision-making performance, supporting the inverted U-shaped relationship between arousal and cognitive performance. By monitoring arousal (e.g., via EmoNet and physiological signals) and adapting explanation delivery accordingly, the system targets the optimal arousal window for effective understanding.
Importantly, the model builds upon grounding theories from human communication (e.g., \cite{brennan1995kbs}), adapting them to a multimodal Human-Agent-Interaction context. 
Future work will evaluate the model’s effectiveness in a controlled user study.


\begin{ack}
This research was funded by the Deutsche Forschungsgemeinschaft
(DFG, German Research Foundation): TRR 318/1 2021-438445824 “Constructing Explainability”.
\end{ack}


\FloatBarrier

\bibliography{References.bib}

\end{document}